\documentclass[referee]{mn2e}
\usepackage{graphics}
\title[The 2175 \AA{\ }-band width]{An enduring puzzle: the width variations of \\
the 2175 \AA{\ } interstellar extinction band}
\author[R. J. Papoular and R. Papoular]{Robert J. Papoular$^{1}$\thanks{E-mail:
Robert.Papoular@cea.fr} and 
Renaud Papoular$^{2}$\thanks{E-mail: papoular@wanadoo.fr}\\
$^{1}$IRAMIS, Laboratoire Leon Brillouin, CEA Saclay, 91191 Gif-s-Yvette, France\\
$^{2}$Service d'Astrophysique and Service de Chimie Moleculaire,
 CEA Saclay, 91191 Gif-s-Yvette, France}
\begin{document}

\date{Accepted . Received ; in original form }

\pagerange{\pageref{firstpage}--\pageref{lastpage}} \pubyear{2002}

   \maketitle
\label{firstpage}

\begin{abstract}
Graphene, a single infinite, planar, sheet of graphite, has the same dielectric resonances as bulk graphite, but solid state theory indicates that its features are about half as wide. Based on this theory,  the dielectric functions of mono- and multi-layer graphenes are  deduced and compared with those of terrestrial graphite. The resonance width of an ordered stack of graphenes is found to increase with the number of layers while the central frequency stays constant. This is the basis of the polycrystalline model of the carrier of the 2175 \AA{\ } interstellar extinction band. In this model, the carrier dust grains derive from parent hydrocarbon grains. As a grain ages in the IS medium, the light atoms are expelled, hexagonal carbon rings lump together into compact planar clusters, which then assemble into stacks of parallel, equidistant, graphene-like layers. This so-called graphitization is well known to occur in the earth or under strong heating. As the number of layers in each stack increases and their relative orientational order improves, the $\pi$ resonance width increases asymptotically towards that of terrestrial graphite. Because of the initial random structure of the parent grains, many randomly oriented stacks may coexist in the same grain. Calculations of the dielectric response of this composite medium show that, for such a grain, the width of the extinction efficiency peak follows the same trend as the $\pi$ resonance of the average stack, and thus covers the observed range of IS feature widths, at very nearly constant peak frequency. 

\end{abstract}





\section{Introduction}

Forty years after its discovery, the 2175 \AA{\ } interstellar absorption band still inspires new efforts in search of a model carrier which could satisfy all the (sometimes apparently contradictory) observational constraints. Among the latest endeavours, we note several papers on fullerenes/buckyonions models (e.g. Wada and Tokunaga \cite{wad} , Chhowalla et al. \cite{chh}, Ruiz et al. \cite{rui}, Li et al.\cite{li}). On the other hand, an improved approach of the polycrystalline graphite (PG) model was described by Papoular and Papoular \cite{pp}.

However, to this day, only a few authors have proposed a solution to the most difficult puzzle in this domain, viz. the relative narrowness of most IS (interstellar) features and the observed variations of their width from 0.7 to 1.3 $\mu$m$^{-1}$ together with the constancy of the central wavelength at 4.6 $\mu$m$^{-1}$ (or 5.7 eV) to better than 2 $\%$. To our knowledge, the latest among them are Sorrell \cite{sor} and Mennella et al. (\cite{men}). Sorrell proposed monosized carbon grains graphitized by UV starlight. Mennella et al. reasoned that an ordered material like graphite is too hard to conceive in space and prefer UV-processed \emph{hydrogenated amorphous ($sp^{2}/sp^{3}$)} carbon grains (BCC: bump carrier carbons). Based on an original approach to their previous spectral reflectance measurements of various samples (Mennella et al. \cite{men96}), they came up with  a series of dielectric functions from which they derived plasmon resonance profiles that fit observations remarkably well.

At about the same time, Duley and Seahra \cite{dul} studied a large number of different theoretical carbon structures similar to those of PAH molecules but with various degrees of hydrogenation, ionization and defects, of which they computed the dielectric functions (using the discrete dipole approximation) and deduced the corresponding bump spectral profiles. While some of these exhibit the bump at the right wavelength, none fulfill the feature width constraints. Again, none exhibit feature widths narrower than 1 $\mu$m$^{-1}$ (their Fig. 16). These authors note that hydrogenation and defects increase the feature width and decrease its intensity, which concurs with previous measurements by Blanco et al. \cite{bla} and Schnaiter et al. \cite{sch96}. As this whole picture seems to be at odds with  Mennella et al.'s favourable conclusion regarding the same type of materials, the issue will be taken up in the Discussion below.

The PG model of Papoular and Papoular \cite{pp} is neither based on the perfectly ordered terrestrial graphite nor on amorphous hydrogenated carbon. It is made of pure, mainly sp$^{2}$ carbon in a disordered structure. Here, based on classical results of solid state physics, as well as on new experimental findings on graphene, we propose that this polycrystalline model can solve the long-standing problem of the IS UV feature width by dint of its specific structure.

The PG model was inspired by reflection measurements on industrial polycrystalline graphite. This is a hydrogen-free  assembly of randomly oriented microscopic sp$^{2}$ carbon chips (or bricks), forming macroscopically homogeneous and isotropic, solid, grains usually delivered in the form of powders of different granulometry, after milling to different extents. In this form, the constituent graphitic chips are usually no smaller than a few tens of nanometers in all three dimensions. They are pressed under more than 10 kbars into disc-shaped pellets. The spectral reflectance $R(\lambda)$ is usually measured at near normal incidence on the circular face of the disc, preferably after some surface smoothing. The Kramers-Kronig (K-K) relations (see Bohren and Huffman \cite{boh}) are then applied to $R$ to deduce the bulk dielectric functions of this material. These are isotropic because, even under the applied pressure, the chips retain their random orientations. 

These experimental functions can be compared with those obtained by applying the Bruggeman mixing formula (see Bohren and Huffman \cite{boh}) to a mixture, in the ratio 1/3 to 2/3, of the dielectric functions of pure, bulk, graphite for $E\parallel c$ and $E\perp c$ orientations, respectively, as determined experimentally. The agreement is adequate provided the powder material contains no volatile atoms such as those that are carried by the binding substances which are sometimes introduced for industrial purposes.

When the \emph{laboratory} PG dielectric functions are used to deduce the extinction efficiency of Rayleigh-sized grains, this is found to have a Fr\"olich (or surface, or plasmon) resonance, nearly Lorentzian in profile, peaking near 4.6 $\mu$m$^{-1}$, with a width FWHM$\sim1.3 \mu$m$^{-1}$. Note that this width falls at the higher end of the spectral interval within which most of the interstellar (IS) feature widths are observed (Fitzpatrick and Massa \cite{fm}, Fitzpatrick \cite{fit}). There is no way of reducing this width, let alone tailoring it, if the bulk graphite properties are retained. Hydrogenation, radiation-induced defects, grain agglomeration are all known to broaden and weaken the plasmon feature.

On the other hand, the recent flurry of research on the electronic properties of graphene provides a treasure trove of experimental facts related to our problem (see Geim and Novoselov \cite{gn}). Graphene is a sheet of compactly assembled hexagonal carbon rings, which can now be produced in the laboratory, free-standing or on a substrate, in nanometric sizes. Multiple layers can also be stacked upon one another in an orderly manner, as in natural graphite (the so-called ABAB, or Bernal, stacking), or in so-called \emph{turbostratic} disorder, i.e. randomly rotated relative to each other, around the normal to the planes, but still parallel and equidistant from each other. The particular finding that hints to a solution of our problem is documented by Ferrari et al. \cite{fer}, who were interested in dc transitions near the Fermi level (the Dirac point, which is the most promising for technological purposes): they observed a continuous increase of the width of the D2 Raman line of orderly stacked sheets of graphene as the number of sheets increases from 1 to 10. This prompted us to enquire if a similar behaviour occurs in the VUV (vacuum ultraviolet) region of the spectrum, near 2175 \AA{\ }. \emph {More precisely, we wish to find out if stacking graphene layers on top of each other can make the Fr\"olich resonance to behave like the IS VUV feature}.

\section{The dielectric function of a single-layer graphene}

Since no laboratory determination of this function is available yet, we have to rely on theory, complemented precisely by the available IS measurements. The imaginary part of the complex dielectric function is (see, for instance, Ziman \cite{zim}, Johnson and Dresselhaus \cite{jd}, Bassani and Pastori-Parravicini \cite{bp})

\begin{equation}
\epsilon_{2}(\omega)\,\propto\,J(\omega)/\omega^{2}
\end{equation}

where the dipole matrix element is assumed constant, and where $J(\omega)$ is the joint density of valence and conduction band states, all over the first Brilloin Zone, that are separated by an energy $\hbar\omega$. As for $\epsilon_{1}$, it is deduced from $\epsilon_{2}$ by the K-K relation. Several approximations to $J(\omega)$ are available in the literature (Johnson and Dresselhaus \cite{jd}, Bassani and Pastori-Parravicini \cite{bp}, Kobayashi and Uemura \cite{kob}, Marinopoulos et al. \cite{mar}, etc.). While their agreement is still not perfect, they are all based on the fundamental van Hove theorem in solid state physics (see Ziman \cite{zim}),which states that, in 2-dimensions (as for planar graphene) and to second order in frequency shift from a resonance frequency, $\omega_{0}$, $J$ behaves like a logarithmic singularity
\begin{equation}
J(\omega)\,\propto\,-{\ln} \!
                            \mid 
                                 1-
                                    {\displaystyle { \omega     }
                                                     \over
                                                   { \omega_{0} }
                                    }
                            \mid + \,\, {\rm constant}
\end{equation}

For reasonable spectral resolutions (e.g. 0.05 eV), the infinitness of $J(\omega_{0})$ presents no problem. However, the far wings have no physical meaning since they reach or cross zero to become negative. So, in order to cover the spectral region of interest, it is necessary to extrapolate the logarithmic wings in a sensible way. Assuming the PG model indeed applies to IS grains, we sought to tailor these wings so as to deliver  the observed range of IS VUV features. For this purpose, we set the additive constant in eq. 2 to zero and multiply by a Gaussian function of width $\gamma$, centered on $\omega_{0}$, and much wider than the logarithmic peak:

\begin{equation}
J(\omega)\,\propto\,-{\ln} \!
                            \mid 
                                 1-
                                    {\displaystyle { \omega     }
                                                     \over
                                                   { \omega_{0} }
                                    }
                            \mid
                     \,{\exp}
                              \left[
                                     - \left(
                                                   \displaystyle{
                                                                  { \omega-\omega_{0}}
                                                                    \over
                                                                  { \gamma           }
                                                                }
                                       \right)^{2}
                              \right]
\end{equation}

Here, $\omega_{0}$ must be set equal to the same $\pi$ or $\sigma$ resonance frequency as in bulk graphite, i.e. 4.3 or 14 eV. This is because the interlayer distance in graphite (3.37 \AA{\ }) is much too large for a layer to notably perturb the resonance of its neighbours, as was recently confirmed by DFT calculations (see Marinopoulos et al. \cite{mar}.

We found that, if the $\sigma$ resonance is omitted, the Fr\"olich feature associated with the $\pi$ resonance is systematically shifted well beyond the observed 5.7 eV (4.6 $\mu$m$^{-1}$). Thus, both $\gamma_{\pi}$ and $\gamma_{\sigma}$ are necessary and sufficient to deliver the desired feature at the right position with the right width. $J(\omega)$ becomes therefore a sum of 2 similar terms.

 In selecting the two $\gamma$ parameters for graphene, we must keep in mind the following: the width of the resonance of a PG-like material made of bricks of stacked graphene layers necessarily exceeds the width of a single graphene resonance, because it is  increased both by stacking and by mixing of $\epsilon$'s for $E\parallel c$ and $E\perp c$, as is shown below. Thus, our scenario only applies if the width of the Fr\"olich resonance of a single layer graphene is found to be at most as wide as the smallest observed IS feature width, namely $\sim0.7\, \mu$m$^{-1}$.

The proportionality factors, multiplying the $\pi$ and $\sigma$ terms in eq. 4 for $\epsilon_{2}$ are determined by application of the Sum Rule, related to the number of electrons per atom contributing to the $\pi$ and $\sigma$ resonances respectively (see Altarelli et al. \cite{as}, eq. 6). In this case, the rule is implemented by requiring that the plateaus of the function 
$\Sigma \epsilon_{2}\omega\,{\rm{d}}\omega$ of $\omega$, at $\sim7$ and $\sim18$ eV respectively, be the same as for graphite as measured by Taft and Philipp \cite{tp}.

\begin{figure}
\resizebox{\hsize}{!}{\includegraphics{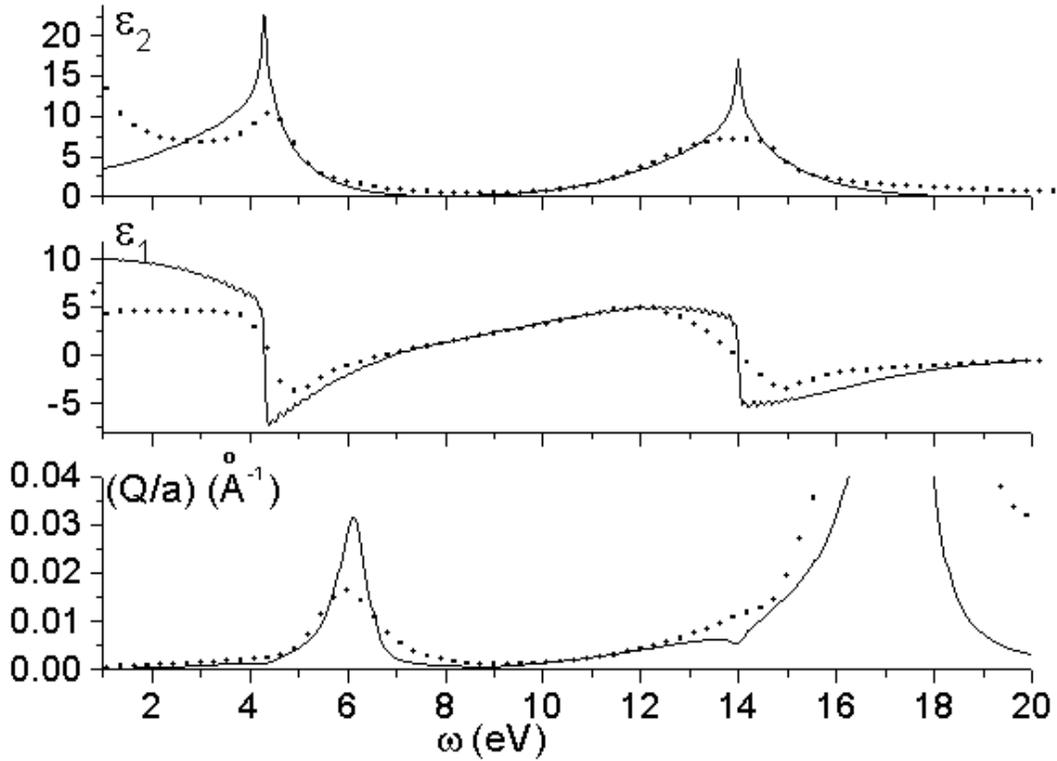}}
\caption[]{Dielectric functions for $E\perp c$; lines for graphene and dots for graphite. From top to bottom: A) $\epsilon_{2}$ for graphene (Sec. 2) and  graphite (Taft and Philipp \cite{tp}); B)$\epsilon_{1}$ for graphene (deduced from $\epsilon_{2}$ by the Kramers-Kronig relation, Sec. 2) and graphite (Taft and Philipp \cite{tp}); C)Extinction efficiency, $Q/a$, for cases graphene and graphite.}
\end{figure}

With the above constraints in mind, we settled on the following expression  for  graphene in the $E\perp c$ polarization;

\begin{equation}
\epsilon_{2}={\displaystyle 1 
                            \over 
                            {{\omega}^{2}}
             }
             \left\{ \!
             A_{\pi}
                       \ln \!
                           \mid 
                           \!
                               1-{\displaystyle { \omega        } 
                                                  \over 
                                                { \omega_{0\pi} }
                                 }
                           \!
                           \mid
                       \exp \!
                           \left[ 
                                \! - \! \left(
                                       {\displaystyle 
                                                       { \omega-\omega_{0\pi} } 
                                                         \over 
                                                       { \gamma_{\pi}         }
                                       }
                                    \right)^{2}
                           \right]
             +
             A_{\sigma}
                       \ln \!
                           \mid
                           \! 
                               1-{\displaystyle { \omega           } 
                                                  \over 
                                                { \omega_{0\sigma} }
                                 }
                           \!
                           \mid
                       \exp \!
                           \left[ 
                                 \! - \! \left(
                                       {\displaystyle 
                                                       { \omega-\omega_{0\sigma} } 
                                                         \over 
                                                       { \gamma_{\sigma}         }
                                       }
                                    \right)^{2}
                           \right]
             \! \right\}
\end{equation}

where $\omega$ is in eV's (1 eV=0.805 $\mu$m$^{-1}$), from 1 to 20 in steps of 0.04 eV, and  $A_{\pi}=-77, \omega_{0\pi}=4.3, \gamma_{\pi}=2.5, A_{\sigma}=-424.3, \omega_{0\sigma}=14, \gamma_{\sigma}=3$.
This equation is plotted in Fig. 1A, together with the corresponding curve for bulk graphite as measured by Taft and Philipp \cite{tp}, for purposes of comparison. The real part of the dielectric function of graphene, $\epsilon_{1}$ was obtained by applying the K-K relation to eq. 4, and is plotted in Fig. 1B, again together with the corresponding function for graphite, as obtained by Taft and Philipp. Finally, the extinction efficiency for both materials was obtained from

\begin{equation}
Q/a=\frac{24 \pi}{\lambda}\frac{\epsilon_{2}}{(\epsilon_{1}+L^{-1}-1)^{2}+(\epsilon_{2})^{2},}
\end{equation}

 where $\lambda$ is the current wavelength and the grain shape parameter is set at $L=1/3$, assuming the grains are roughly spherical. Only the $\pi$ resonance part of $Q/a$, which is of interest for comparison with IS spectra, is plotted in Fig. 1C, for graphene (the narrowest peak, 0.75 eV or 0.6 $\mu$m$^{-1}$) and graphite (the widest peak, 1.5 eV or 1.2 $\mu$m$^{-1}$).

Admittedly, the Gaussian function was introduced in eq. 3 as an \emph{ansatz} to obtain the desired feature at the right location with the right width and the right intensity. However, since this IS feature is so sensitive to the dielectric function in its spectral range, and if our graphene model is indeed valid, our procedure can also be considered as a guide to the theory of graphene, based on accurate astronomical measurements. This is welcome, since even \emph{ab initio} theories of the joint density of states must rely on empirical coupling energies and, moreover, hardly agree with each other (cf. for instance Bassani and Pastori-Parravicini \cite{bp}, and Marinopoulos et al. \cite{mar}).

Besides, the use of Gaussians is not new in solid state spectroscopy. For instance, Dasgupta et al. \cite{das} used a Gaussian multiplied by an error function to fit the absorption of a-C and a-C:H in their band gap. Kim et al. \cite{kim} defined a combination of Gaussian and Lorentzian profiles (so-called pseudo-Gaussian) to model the optical dielectric functions of semiconductors and Djurisic et al. \cite{dju} applied the same device to the case of graphite. In atomic spectroscopy, another combination of Lorentzian and Gaussian is often used: the Voigt profile. In general, Gaussians are helpful in representing resonance wings.

\section{From graphene to graphite}

What are the intermediate steps between graphene and graphite? Our model assumes that, during their sojourn in the interstellar medium, carbon-rich grains are initially in the form of hydrocarbon chains, benzenic rings and aromatic clusters of various sizes. Under the combined effects of more or less strong irradiation and shocks, they progressively lose their volatile atoms
(essentially hydrogen) at correspondingly higher or lower rates. The unavoidable consequence of this ``carbonization" and of the chemical nature of carbon is that the remaining chains bend into rings, the rings are rearranged into more compact flakes and the flakes tend to pile up in stacks (``graphitization"). The whole process is well documented in the earth (see Durand \cite{dur}; Bustin et al. \cite{bus}) where, under high pressure and temperature, it can give birth to perfectly graphitized crystals of up to 1 in. By contrast, in the interstellar medium, the graphitization process is limited by the low ambient density and, moreover, is interrupted by grain collisions, giving birth, instead, to what we believe looks like laboratory polycrystalline graphite.

Let us first assume that the size of each graphene layer in an IS grain is large enough that the dielectric function obtained above for infinite planes also applies here (this will be discussed below). As in the laboratory (see above), the stacking of successive layers can occur in an orderly (ABAB) manner or in turbostratic disorder. If the latter is the case, then it is reasonable to assume that the bulk dielectric function of a brick is some average of the dielectric functions of graphene and graphite, weighted according to the degree of disorder. Hass et al. \cite{has} showed that for complete disorder, the stack behaves like a single sheet of graphene.

If, on the other hand, the stacking is orderly, then the situation can be visualized by considering each layer as a harmonic oscillator resonating at one of the $\pi$ electron frequencies (the analog of quantum states) forming a band. If one identical layer is added and is electrically coupled to the previous one, then the frequency degeneracy is lifted by the perturbation, as is well known, and two new resonance frequencies appear instead, bracketing the original one, and separated by an interval $\Delta\omega$ proportional to the coupling coefficient between two adjacent layers, and roughly equal to the FWHM of the $\pi$ resonance, which is $\sim1.5$ eV for graphite (Bassani and Pastori-Parravicini \cite{bp}, Marinopoulos et al. \cite{mar}, Klintenberg et al. \cite{kli}). For N layers, there are N new resonance frequencies  (or electronic quantum states) but these are all included in this interval (see Gr\"uneis et al. \cite{gru}, Fig. 13). In graphite, N is very large and the interval is densely filled. If the full natural width of each resonance is $\delta\omega$ (that of a single layer graphene, which we found in Sec. 2 to be $\sim0.75$ eV) it is apparent that the minimum number of oscillators (i.e. graphene layers) necessary to practically ``fill" that interval is N$_{m}=2\,\,\Delta\omega/\delta\omega$. For the widths adopted above, this turns out to be $\sim4$. Many measured properties of graphite are found to be practically recovered by stacking about 5 graphene layers (see Geim and Novoselov \cite{gn}, Michel and Verberck \cite{mv}, Fig. 6).

 This must hold for the dielectric functions as well: as the number of layers increases beyond 1 and the number of quantum states that lie within $\Delta\omega$ increases proportionately, the 3-dimensional character of the stack increases linearly with N. For a stack of N layers, if we define a relative number of layers as $s$=(N-1)/N$_{m}$, a resonable approximation to the average dielectric function is, therefore,

\begin{equation}
\epsilon(\rm{bulk})=\frac{\epsilon(\rm{graphene})+s\,\epsilon(\rm{graphite})}{1+s}.
\end{equation}

However, this has also to be averaged over all bricks in a grain, so N is replaced by its average $\bar{\rm{N}}$, and we end up again with a weighted average of the dielectric functions of graphene and graphite. We assume this is true for both polarizations of the electric field.

\begin{figure}
\resizebox{\hsize}{!}{\includegraphics{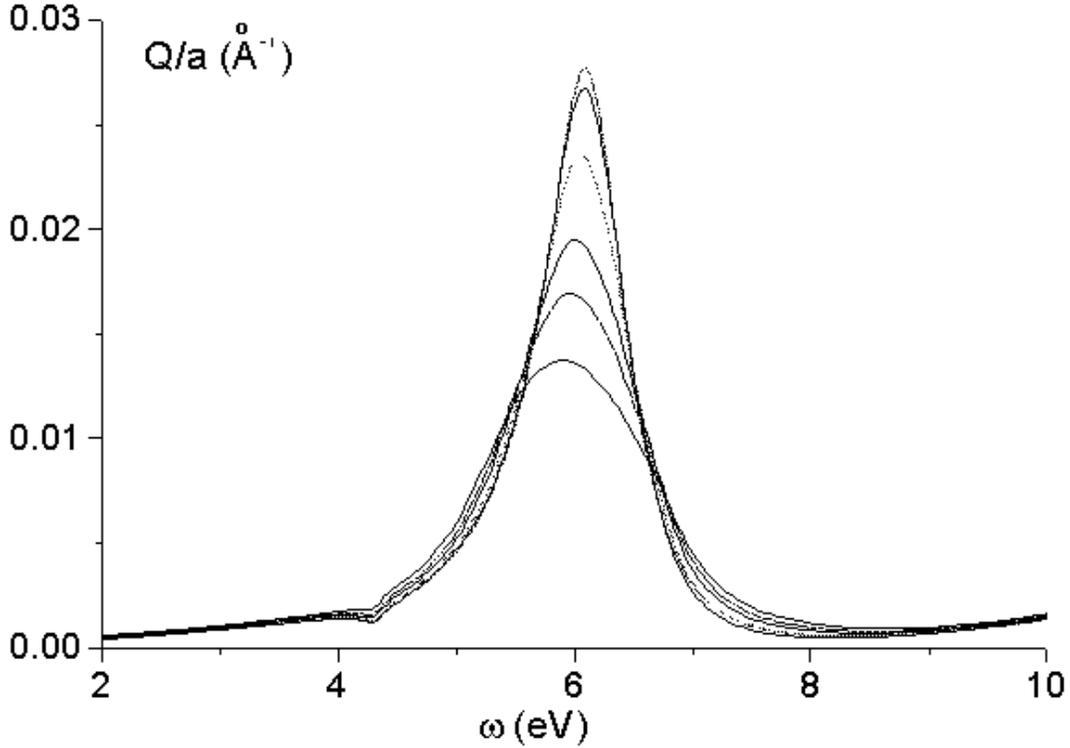}}
\caption[]{Evolution of the (Fr\"ohlich) surface-resonance associated with the $\pi$ resonance of Rayleigh-sized polycrystalline grains as the average relative number, $s$, of graphene layers per stack (constitutive brick) increases. From the strongest to the weakest peak, $s=0$ (graphene), 0.1, 0.2, 0.4, 0.6, 1 (quasi graphite). From $s=0.1$ to 0.8, the feature width increases from 0.9 to 1.6 eV ($\sim0.73$ to 1.29 $\mu$m$^{-1}$), while the peak shifts from 6.1 to 6 eV ($\sim4.9$ to 4.8 $\mu$m$^{-1}$).}
\end{figure}

Remembering that in a given model IS grain, the graphitic bricks are randomly oriented, we have now to apply the Bruggeman mixing formula to mixtures of parallel and transverse polarizations, as for PG (see Introduction). The dielectric function of graphite in the parallel polarization is subject to great uncertainty because of the experimental difficulty of presenting a proper edge face to the light beam (see discussion in Draine and Lee \cite{dl}). A further uncertainty is introduced by the effects of the number of layers in a stack: in preliminary studies, the dc conductivity perpendicular to the layer plane was shown to increase distinctly with the number of layers (see Fei et al. \cite{fei}). Assuming this is also true in the optical range, and by analogy with our treatment of $\epsilon\perp$, we shall therefore take $\epsilon\parallel$ to be similarly proportional to $s$, which would entail a change of the mixing ratio, $f$, in the Bruggeman formula from 1/3 to $s/3$ (to a factor), ensuring that the contribution of the parallel polarization is null for graphene. In the absence of consensual laboratory measurements of $\epsilon\parallel$ for multi-layer graphene, we let ourselves be guided again by the IS feature measurements: we adopt the dielectric function tabulated by Draine \cite{dra}, and take $f=s/6$, which gives a better fit to observations. Figure 2 shows the extinction feature for $s=0$ (i.e. graphene), 0.1, 0.2, 0.4, 0.6 and 1 (i.e. close to graphite). Clearly, the extinction feature can be controled through $s$, i.e. the average number of layers per stack in a grain while the peak frequency shift remains within the observational limits. Note, however, that this frequency is 6 $\%$ higher than the observed 5.7 eV (4.6 $\mu$m$^{-1}$). But this is within the experimental errors affecting the measurements of $\epsilon_{2}$(graphite): its $\pi$ peak is found at $\sim$4.3 and $\sim$4 eV by Taft and Philipp \cite{tp} and Tosatti and Bassani \cite{tb}, respectively. Had we adopted the intermediate value 4.1, in Sec. 2, the peaks in Fig. 2 would have been shifted to the IS position.

Thus, our model is able to satisfy the simultaneous constraints of variable feature width and nearly constant peak position. This is effected by changing only the number of graphene layers per stack, which is positively correlated with the degree of graphitization of the stacks in the grain, and hence to the amount of processing of the dust by UV heating. Figure 1 illustrates the change of properties encompassed by this model, between its extremes: graphene and graphite. Note that the model does not specify the number of stacks per grain. A grain could well consist of only one stack (brick). 

While this model exhibits a far UV rise in extinction, it should be stressed that possible continuum contributions to extinction by other carriers are outside the scope of this work.

\section{Discussion}
\subsection{The size of the graphene sheets}
The argument developed above hinges upon the graphene properties. However, graphene theory assumes an infinite plane. Can this be applied to our model graphitic bricks? The latter must be much smaller than the average grain size, which is itself limited to $\sim300$ \AA{\ } by the Rayleigh condition for 2175 \AA{\ }. We have to enquire if the properties of graphene still apply at smaller sizes.

These properties are determined by the interactions between an atom and its neighbours. Obviously, these interactions fade away as the distance to the central atom increases. 
In order to find the ``cut-off" distance, we therefore considered successively compact clusters of carbon rings of increasing size. With the help of a commercial quantum chemistry package (Hypercube, Hyperchem 7), their structure was optimized by seeking the minimum total binding energy,  using the semi-empirical AM1/UHF method. Figure 3 shows the variation of the binding energy per atom, $e_{b}$, as a function of the number of rings, N$_{\rm{r}}$. 

\begin{figure}
\resizebox{\hsize}{!}{\includegraphics{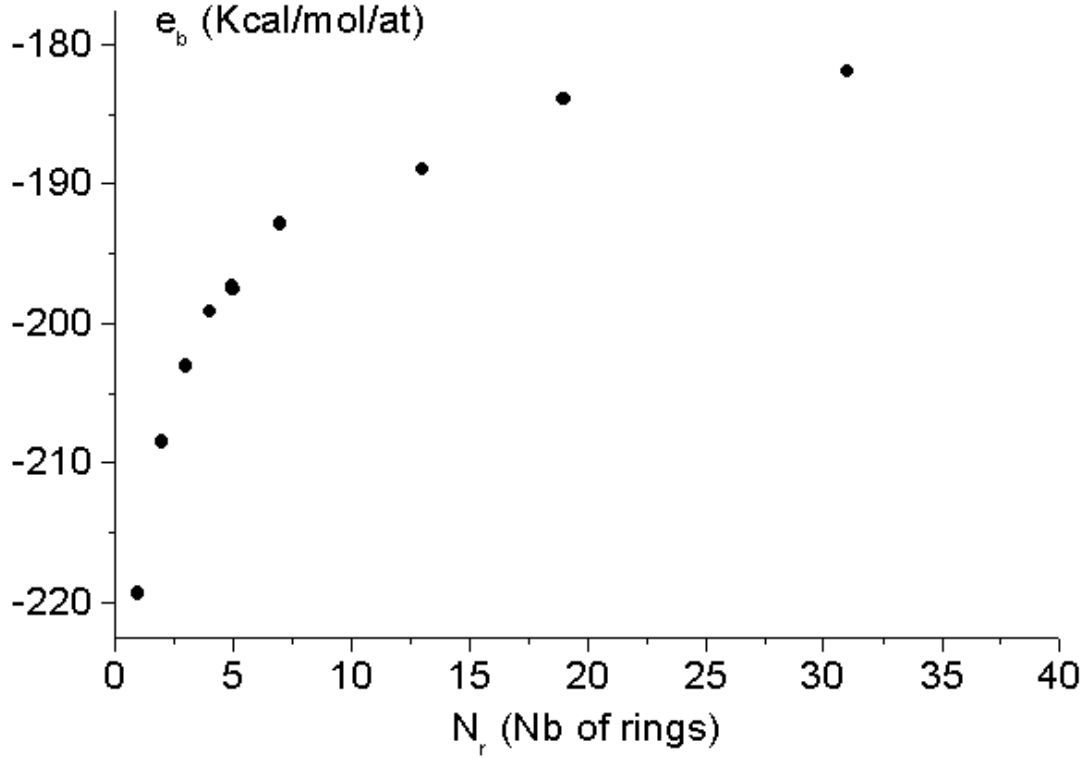}}
\caption[]{Binding energy per atom (Kcal/mol) as a function of the number of benzenic carbon rings in planar, compact clusters; 1 Kcal/mol=0.0434 eV.}
\end{figure}

Beyond about 40 rings, or 84 atoms, the binding energy reaches, for all practical purposes, an asymptotic value of 182 kcal/mol or 8 eV per atom, which is indeed about the graphite sublimation energy. The corresponding diameter of the structure is about 1.5 nm. A similar conclusion was reached by Robertson \cite{rob} on the basis of a simpler H\"uckel calculation. Thus, the minimum extension required for a graphene sheet is small enough that many such sheets can be accommodated in a Rayleigh-sized grain.

The structure of a fraction of the IS grains has certainly evolved to the point where the dielectric properties of bulk graphite apply, as witnessed by the fact that the maximum observed feature width is accounted for by the measured graphite dielectric functions (see Sec. 1). It cannot be excluded, however, that another family of IS grains harbors smaller graphene sheets. Should that be the case, the effect on the $\pi$ resonance could be approximated, to first order, by a reduction of the electronic lifetime and hence, a widening of the band. For a sheet size  $L_{a}$ and an electronic velocity of $5\,\,10^{7}$ cm.s$^{-1}$ (see Williamson et al. \cite{wil}), the contribution of brick size to the widening of the $\pi$ resonance is, therefore, 
\begin{equation}
\Delta\Gamma (\mu \rm{m}^{-1})=\frac{1}{2\pi\tau c}=\frac{2.5}{L_{a}(\rm{\AA{\ }})},
\end{equation}
 where $c$ is the velocity of light and $\tau$ the time of flight of an electron across a sheet. This is about 0.1 $\mu$m$^{-1}$ for the minimum sheet size and decreases as the size increases. As shown by Papoular and Papoular \cite{pp}, the Fr\"ohlich resonance is consequently widened by the same amount. The contribution of this effect cannot be dominant, for otherwise the maximum observed extinction band width would be larger. However, it may not be negligible for lesser evolved grains with a small number of sheets per stack.

Note that the electronic band theory of solids was used above for the 2-D graphene, but not for graphite, for which we chose the firmer ground of laboratory measurement. Similar theories exist however for the 3-D case and have been applied to graphite (see Coulson and Taylor \cite{ct}, Kobayashi and Uemura \cite{kob}). Their results confirm the widening of the resonance bands as more graphene sheets are orderly stacked upon one another, and support our calculations with further \emph{ab initio} arguments.

\subsection{Relation to amorphous carbons}
It has become clear over the years that, although carbon may be the prime contender as carrier of the IS UV feature, the initially proposed perfect graphite embodiment does not completely satisfy the astronomical constraints. As a consequence, various related avenues were explored. We have briefly discussed the fullerenes and bucky onions (Papoular and Papoular \cite{pp}), of which both measurements and theoretical computations show an excellent fit to the broadest IS features. Here, we turn to the other extreme, namely amorphous carbon.

It is hardly possible to cite exhaustively here all the results published in this field, so we shall be content with referring to the proceedings of the Capri Conference (Bussoletti et al. Eds. \cite{bus}) where many  contending models were described. Essentially, the experimental procedure consists in producing a hydrocarbon vapour, either by combustion of a hydrocarbon gas or by an electrical discharge between amorphous carbon electrodes in vacuum or inert gas. The carbon vapour is then made to condense homogeneously on fused silica substrates, to form a soot composed of solid grains. These grains may later be treated by heating or irradiation (UV or ions). The situation in 1996 was summarized by one of the most active team (Blanco et al. \cite{bla}), who also compared all other carbon models studied at the time. On this basis, it seems fair to define the consensus as follows: as deposited, the resulting materials exhibit weak or no UV feature. When treated by heat or UV, they deliver a feature which progressively increases in intensity and central wavelength, and decreases in width. This is associated with loss of hydrogen and growth of aromatic clusters. Partial processing is able to bring the central wavelength at the right value ({\ }1275 \AA{\ }), but it is seems unlikely that such specific processing could occur systematically in space. Blanco et al. proposed instead the inclusion, in their grains, of crystallites of a still more graphitic material, like certain types of glassy carbon. They noted the similarity of this model with the polycrystalline graphite model of Papoular et al. \cite{pap93}, and rightly concluded that none of the available models delivered sufficiently narrow UV feature to fit IS observations. However, they suggested that clustering of primary grains produced in the laboratory could contribute to that excessive feature width.

Independently, Schnaiter et al. \cite{sch96} produced their amorphous carbon grains in a flame, by burning acetylene with oxygen at low pressure. They prevented their coagulation by isolating them in a icy argon matrix, thanks to elaborate beam techniques. They obtained independant primary particles, 6 nm in diameter on average, of which they measured the UV extinction. Their feature peaked at $\sim240$ nm, with a width of 54 nm ($\sim1 \mu$m$^{-1}$). This width is probably the narrowest ever obtained with any laboratory material; however, it is not clear how it can be varied about the measured value to fit the observed range (0.7 to 1.3 $\mu$m$^{-1}$). Moreover, the wavelength of the corresponding peak is distinctly redshifted relative to the IS value.
 
These experimental results were backed by the theoretical computation of the dielectric properties of many different carbon structures by Duley and Seahra \cite{dul}. As noted in the Introduction, these authors do not consider hydrogenated amorphous carbons to be the ultimate solution to our conundrum. It remains therefore to understand the apparent contradiction with the computations of Mennella et al.\cite{men} referred to in the Introduction.

The latter computations were based on their previous transmission measurements through UV irradiated hydrogenated amorphous carbon (HAC) samples (Mennella et al. \cite{men96}), which delivered UV extinction peaks quite akin to those of Blanco et al. \cite{bla}, except that they do not reach shortward of 190 nm. Their new approach to the experimental results consists essentially in assuming that the apparent discrepancies with IS observations are due to the adverse effects of agglomeration of primary grains produced in the laboratory, into continuous distributions of ellipsoids (CDE). This procedure is questionable, as we show below.

First, note that the extinction spectra of their UV irradiated samples (to which they fit the CDE model) are limited to photon energies weaker than 6.5 eV (their Fig. 1), and the dielectric functions they deduce therefrom are shown up to only 5.3 $\mu$m$^{-1}$, i.e. still within the bump width and not even including the $\pi$ resonance (Fig. 2). This gives a very restricted and baffling  picture of the UV spectrum of the materials: it would be highly desirable to show both the $\pi$ and $\sigma$ resonance peaks, which must be there, and are good indicators of the carbon $sp^{2}$ character. The $\sigma$ resonance must be quite strong, judging from the non-irradiated samples in Fig. 1. Its complete neglect in this treatment is all the more worrying that our target feature falls in between the two resonances, and is critically sensitive to variations of the slopes of both the real and imaginary dielectric functions in that interval, which are themselves highly dependant on both resonances.

Furthermore, no clear experimental evidence is given for the use of the CDE paradigm in the present case, instead of assuming for instance that what is measured is the extinction through the bulk primary material or specific clusters as imaged by Rotundi et al. \cite{rot}. Such a justification is essential since a combination of shapes ranging from needles to discs is known to produce a huge artificial broadening of primary resonances (see Schnaiter et al. \cite{sch98}). So the inferred resonance width for \emph{spherical} grains, made of the primary material, is obviously much narrower than suggested by the ``raw" spectrum of the sample. To avoid pitfalls one had better compress the material into dense pellets and measure their reflectance, or isolate the primary grains and measure their extinction, as did Schnaiter et al. \cite{sch96} and Schnaiter et al. \cite{sch98}.

Finally, the fits to the narrowest IS features use dielectric functions which were not inferred from measurements on the two irradiated samples, but obtained by ``linear extrapolation" of the latter, for the higher radiation doses which could not be applied in the laboratory (Fig. 2, BCC I to IV). It is not clear how whole dielectric function spectra can safely be extrapolated from two extreme ones (BCC V and VI), which fall at the same edge of the range.

In conclusion, it appears to us that the best picture of the present performance of amorphous carbon, as a model for the IS feature, for comparison with other models, remains that given by Blanco et al. \cite{bla} and Schnaiter et al. \cite{sch98}.

\section{Acknowledgments}
We gratefully acknowledge fruitful discussions with, and precious information from, Drs F. Mauri and M. Calandra (IMPMC, Paris), Dr V. Olevano (Inst. Neel, Grenoble, France) and Dr L. Henrard (Univ. N.-D. de la Paix, Namur, Belgium).

\end{document}